\documentclass[12pt,preprint]{aastex}

\shorttitle{Simulating {\it Astro-E2} Observations}
\shortauthors{Fujita et al.}

\begin{document}

\title{Simulating {\it Astro-E2} Observations of Galaxy Clusters: the
Case of Turbulent Cores Affected by Tsunamis}

\author{Yutaka Fujita\altaffilmark{1,2}, Tomoaki
Matsumoto\altaffilmark{3}, Keiichi Wada\altaffilmark{1,2}, and Tae
Furusho\altaffilmark{4}}

\altaffiltext{1}{National Astronomical Observatory, Osawa 2-21-1,
Mitaka, Tokyo 181-8588, Japan; yfujita@th.nao.ac.jp}
\email{yfujita@th.nao.ac.jp}

\altaffiltext{2}{Department
of Astronomical Science, The Graduate University for Advanced Studies,
Osawa 2-21-1, Mitaka, Tokyo 181-8588, Japan}

\altaffiltext{3}{Department of Humanity and Environment, Hosei
University, Fujimi, Chiyoda-ku, Tokyo 102-8160, Japan;
matsu@i.hosei.ac.jp}

\altaffiltext{4}{Institute of Space and Astronautical Science, Japan
Aerospace Exploration Agency, 3-1-1 Yoshinodai, Sagamihara, Kanagawa
229-8510, Japan; furusho@astro.isas.jaxa.jp}

\begin{abstract}
This is the first attempt to construct detailed X-ray spectra of
clusters of galaxies from the results of high-resolution hydrodynamic
simulations and simulate X-ray observations in order to study velocity
fields of the intracluster medium (ICM). The hydrodynamic simulations
are based on the recently proposed tsunami model, in which cluster cores
are affected by bulk motions of the ICM and turbulence is produced.  We
note that most other solutions of the cooling flow problem also involve
the generation of turbulence in cluster cores.  From the mock X-ray
observations with {\it Astro-E2 XRS}, we find that turbulent motion of
the ICM in cluster cores could be detected with the satellite. The
Doppler shifts of the metal lines could be used to discriminate among
turbulence models.  The gas velocities measured through the mock
observations are consistent with the line-emission weighted values
inferred directly from hydrodynamic simulations.
\end{abstract}


\section{Introduction}

The advent of {\it Astro-E2} \citep{ino03b} will enable us to directly
measure velocity fields of the intracluster medium (ICM) in clusters of
galaxies for the first time. The calorimeter ({\it X-Ray Spectrometer;
XRS}) of {\it Astro-E2} has an excellent spectroscopic resolving power
\citep{kel04,cot04}, and it could detect the bulk gas motion by
observing the energy shift of metal lines and the turbulence by
observing broadened metal lines. On the other hand, many hydrodynamic
simulations have been performed to study the motion of the ICM
\citep[e.g.][]{evr90,tak99,yos00,ric01}. With {\it Astro-E2}, we could
confirm the results of those hydrodynamic simulations. However, the
direct comparison of X-ray observations to hydrodynamic simulations is
not always simple, because of complex such as instrumental responses.

In this paper, we construct the X-ray spectra from the results of
hydrodynamic simulations of cluster cores affected by `tsunamis'
\citep*[][hereafter Paper~I, see also \citealt*{fuj04b}]{fuj04a} and we
simulate observations with {\it Astro-E2 XRS}. In the tsunami model,
large-scale bulk gas motions in the ICM, which are called tsunamis,
induce fully-developed turbulence in cluster cores. The ultimate
spectroscopic capability of {\it XRS} with an energy resolution of $\sim
6$~eV at 6~keV for extended sources as well as point sources could give
us a new approach to detect the turbulence in the ICM. Because of the
kinetic energy of turbulence and the hot gas brought into the core by
turbulent mixing, the radiative heating of the core is suppressed. Thus,
this model could solve the so-called `cooling flow problem'
\citep{mak01,pet01,kaa01,tam01}.

From the mock observations with {\it Astro-E2}, we measure velocity
fields of the ICM and compare them with the emission weighted values
inferred directly from the hydrodynamic simulations. We investigate
whether the former is consistent with the latter. Most observations of
clusters scheduled as {\it Astro-E2} performance verification targets in
the first 6 months are planned pointing at their
centers\footnote{http://www.astro.isas.jaxa.jp/astroe/proposal/swg/swg\_lst.html};
one of the reasons is that the effective area of the {\it XRS} is
relatively small and objects must be bright enough to be observed. Thus,
it would be useful to simulate the observations focused on cluster
cores. It should be noted that other solutions of the cooling flow
problem besides the tsunamis, especially those based on AGN activities,
also predict similar level of turbulence in cluster cores, which some
observations have already suggested
\cite[e.g,][]{bla01,chu02,bru02,kim03,kai03,fab03,sok04}.
%
%Recently, \citet{gar04} constructed the software package X-MAS, a tool
%devoted to simulate X-ray observations of galaxy clusters obtained from
%hydro-{\it N}-body simulations. However, they focused on creating X-ray
%maps and did not treat the velocity fields of the ICM. As far as we
%know, there have been no studies to construct detailed X-ray spectra of
%clusters from the results of hydrodynamic simulations and simulate X-ray
%observations in order to study the velocity fields of the ICM. 
In this paper, we assume that $\Omega_0=0.27$, $\lambda=0.73$, and
$H_0=70\rm\; km\; s^{-1}\; Mpc^{-1}$.

\section{Construction of X-Ray Spectra}
\label{sec:const}

We use the results of hydrodynamic simulations done in Paper~I.  These
simulations are performed using a two-dimensional (cylindrically
symmetric) nested grid code \citep{mat03}, and the coordinates are
represented by $(R,z)$.  Since the structure of turbulence in the
axi-symmetric coordinate might be different from that in the fully
three-dimensional coordinate, it is ideal to perform three-dimensional
calculations.  However, a high spatial resolution with a large dynamic
range is also crucial to simulate a turbulent medium, especially for the
tsunami model. Therefore we here restrict the calculations to two
dimensions. The cluster center corresponds to $(R,z)=(0,0)$. Although
the maximum resolution of the simulations is achieved on the level of
grids of 22~pc, we use the simulation results on the level of the grids
of 1.4~kpc, because the turbulent velocity is generally smaller on
smaller scales and it helps us reduce the number of computational grid
points for which we calculate the spectra. We will argue that the effect
of the lower resolution on the X-ray spectra can be ignored in
\S\ref{sec:dis}. Note that in the nested grid code we used, calculations
are performed on all grid levels simultaneously. At the cluster center,
for example, the solutions (density, temperature, etc.) are obtained on
the seven different resolutions. The results of the lower resolution
grids are just the coarsened ones of the higher resolution grids.

We use the simulation results for $256\times 512$ computational grid
points. For each grid point, we calculate the X-ray spectrum using a
single {\tt bapec} thermal model and the data simulation command {\tt
fakeit} in the XSPEC package (version~11.3.1). The {\tt bapec} model is
the same as the {\tt apec} model but includes a parameter for turbulent
velocity broadening to emission lines. The input parameters of the
spectrum for the $i$-th grid point are the temperature ($T_i$), the
metal abundance ($Z_i$), the gas velocity along the line of sight
($v_i$), and the normalization. The turbulent velocity for each grid
point is assumed to be zero, that is, the gas velocity is uniform within
a single grid point. We used the {\it Astro-E2 XRS} response file ({\tt
xrs\_ao1.rmf}) and the auxiliary file ({\tt xrs\_onaxis\_open\_ao1.arf})
provided by the {\it Astro-E2} team as planning tools for AO-1
proposals\footnote{http://www.astro.isas.jaxa.jp/astroe/proposal/ao1/rsp/index.html.en}.
Weighting the three-dimensional volumes of the grid points, the spectra
are summed up by the FTOOLS manipulation task {\tt mathpha}.  In order
to avoid the error caused by small photon counts, we multiply the photon
counts for each grid point by 100000. After the summation by {\tt
mathpha}, the total photon counts are divided by 100000 using the {\tt
fcalc} task in the FTOOLS. Since {\tt fcalc} discard fractions in the
calculation of the photon counts, we add 50000 to the counts before the
division in order to round off. For the resultant spectrum, the error of
photon counts in each energy bin is assumed to be the square root of the
photon counts in the bin. Background emission and Galactic absorption
are ignored. We do not consider the effect of resonance scattering for
metal emission lines, because the calculation of multi-dimensional
radiative transfer is required and it is beyond the scope of this paper.

\section{Results} 
\label{sec:results}

In paper~I, we calculated the evolution of the cool core of a cluster
for $R\lesssim 300$~kpc and $|z|\lesssim 300$~kpc. We approximated the
bulk gas motions in a cluster by plane wave-like velocity perturbations
in the $z$-direction represented by $\delta v_z=\alpha c_s \sin(2\pi c_s
t/\lambda)$ at $z=-345$~kpc, where $c_s$ is the initial sound velocity,
and $\lambda$ is the wave length. The waves propagate in the
$z$-direction. At $t=0$, the cluster is isothermal ($T=7$~keV). Metal
abundance is uniform and $Z=0.3\: Z_\sun$.  We construct X-ray spectra
for two models with different $\lambda$ (Table~\ref{tab:model}). The
model of $\alpha=0.3$ and $\lambda=100$~kpc (Model~A) was studied in
Paper~I. The model of $\alpha=0.3$ and $\lambda=1000$~kpc (Model~B) is
newly studied and the detailed analysis of the results will be discussed
elsewhere (Matsumoto, Fujita, \& Wada 2004, in preparation). Compared to
Model~A, Rayleigh-Taylor and Kelvin-Helmholtz instabilities develop on
larger scales in Model~B. This makes turbulent mixing and heating more
effective.

We assume that the model cluster is at redshifts of 0.01, 0.04,
and~0.08, although the temperature outside of the core is $\sim 7$~keV
and there is no such a high-temperature cluster at redshift of
0.01. Since the field of view of {\it Astro-E2 XRS} is $2.9'\times
2.9'$, we consider the X-ray emission within a radius of $1.5'$ from the
cluster center. At redshifts of 0.01, 0.04, and~0.08, the angle of
$1.5'$ corresponds to 18.5, 71.3, and~136~kpc, respectively. We
`observe' the model cluster along the $z$-axis, that is, the line of
sight is assumed to be parallel to the $z$-axis. This is because the
waves were injected along that axis, and thus the ICM velocity in the
$z$-direction is much larger than that in the $R$-direction. Thus, we
sum up the spectra of the ICM in individual grid points for $R<18.5$,
71.3, and~136~kpc and $|z|<300$~kpc for the cluster at redshifts of
0.01, 0.04, and~0.08, respectively. This means that for a cluster at a
larger redshift, we observe a larger area. The assumed exposure time is
50~ks. Using XSPEC, we fit the summed spectra with a single {\tt bapec}
model. The spectra were grouped to have a minimum of 50 counts per
bin. We limited the energy range to 5--10~keV (at the cluster-rest
frames) that includes Fe--K lines ($\sim 6.7$~keV). Free parameters in
the {\tt bapec} model are the temperature ($T$), the metal abundance
($Z$), the average velocity ($v_{z,sp}$), the turbulent velocity
($\sigma_{z,sp}$), and the normalization. We choose $t=3.3$~Gyr for
Model~A; the temperature distribution at that time is shown in
Paper~I. We choose $t=5.0$~Gyr for Model~B.\footnote{Movies are
available at http://th.nao.ac.jp/tsunami/index.htm .}
%The temperature
%distribution at that time is shown in Figure~\ref{fig:temp}. 
For Model~A, the gas temperature in at least one of the grid points
reaches zero at $t=3.3$~Gyr (Paper~I), while for Model~B, it reaches
zero at $t=6.2$~Gyr. Note that the time-scale of 6.2~Gyr is comparable
to the typical age of clusters \citep[e.g.][]{kit96}, and thus radiative
cooling is almost completely suppressed in Model~B.

\begin{figure}
\epsscale{.50} \plotone{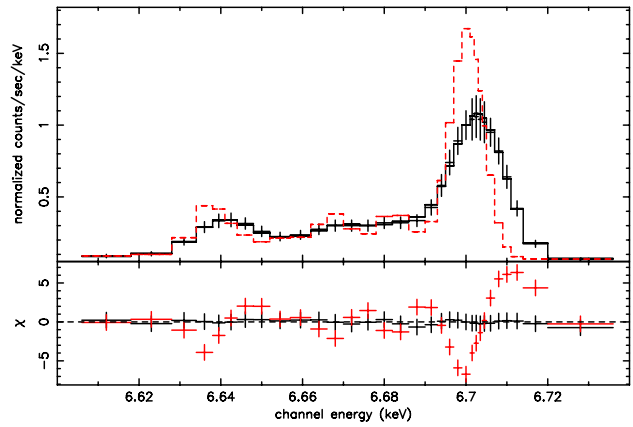} \caption{Upper panel shows the simulated
X-ray spectrum observed with {\it Astro-E2 XRS} for Model~B at redshift
of 0.04 (crosses) and the best-fit model (solid line). The spectrum when
$v_{z,sp}=0$ and $\sigma_{v,sp}=0$ is also shown (dashed line). The
spectra are shown at the cluster-rest frame. The lower panel plots the
residuals divided by the $1\sigma$ errors.}  \label{fig:sp}
\end{figure}

Figure~\ref{fig:sp} shows the simulated X-ray spectrum around $6.7$~keV
Fe--K lines for Model~B at redshift of 0.04 and the result of the
fit. For comparison, the spectrum when $v_{z,sp}=0$ and
$\sigma_{z,sp}=0$ is also shown. As can be seen, the two spectra
(i.e. with and without turbulent motion) are remarkably different.  The
results of the spectral fits for all models are shown in
Table~\ref{tab:para}. Errors on fitted spectral parameters are given at
the 90\% confidence level. In Figure~\ref{fig:vel}, we show $v_{z,sp}$
and $\sigma_{z,sp}$. Most models require non-zero turbulent
velocity. The values of $\chi^2$ are generally small and the fits are
good. The results show that the line shift and turbulence will be
detected with {\it Astro-E2 XRS} if $v_{z,sp}\gtrsim 100\:\rm km\:
s^{-1}$ and $\sigma_{z,sp}\gtrsim 100\:\rm km\: s^{-1}$, because
$v_{z,sp}=0$ and $\sigma_{z,sp}=0$ are rejected, respectively
(Table~\ref{tab:para}). However, it should be noted that our results
correspond to an ideal case. For example, we did not consider possible
systematic errors remaining even after calibrations. Moreover, if the
line of sight is perpendicular to the $z$-axis, observed average and
turbulent velocities are smaller. Thus, turbulence could not be detected
for all clusters.

For comparison, we present the luminosity-weighted turbulent velocities
in the $z$-direction that are directly derived from the results of
hydrodynamic simulations. They are given by
\begin{equation}
\label{eq:sigma_zl}
 \sigma_{z,l} \equiv \sqrt{\frac{\sum_{i} (v_{z,i}-v_{z,l})^2 L_i}
{\sum_i L_i }} \;,
\end{equation}
where $v_{z,i}$ is the gas velocity in the $z$-direction, and $L_i$ is
the luminosity of the $i$-th grid point. As for the luminosity, we
consider both the bolometric luminosity, $L_{bol,i}$, and the Fe--K-line
luminosity $L_{Fe,i}$. Using XSPEC and the {\tt bapec} model, we found
that the emissivity to be used for the latter is approximately given by
\begin{eqnarray}
 \epsilon_{Fe} &=& 5.559\times 10^{-24} n_e n_{\rm H} Z\:\: 
  {\rm ergs\: cm^{-3} s^{-1}} \\ \nonumber
& & \times
 \left[\frac{25.48\:(0.2917 + T)^2}{13.14 + (0.00169 + T)^2}
+1-4.215\: T + 0.2492\: T^2 - 0.005255\: T^3\right] \: 
,
\end{eqnarray}
for the 6.6--6.72~keV band, where $T$ is the ICM temperature (keV), $Z$
is the metal abundance ($Z_\sun$), and $n_e$ and $n_{\rm H}$ are the
electron density and the hydrogen density ($\rm cm^{-3}$),
respectively. This approximation holds within a few percent for $1.5\leq
T \leq 15$~keV (even at $T\sim 1$~keV, the error is $\lesssim
10$~\%). We ignore the Fe--K emission from the ICM of $T<1.5$~keV,
because $\epsilon_{Fe}\approx 0$ and the mass of such gas is small in
our calculations. In equation~(\ref{eq:sigma_zl}), the average gas
velocity in the $z$-direction, $v_{z,l}$, is given by
\begin{equation}
 v_{z,l} \equiv \frac{\sum_{i} v_{z,i} L_i}
{\sum_i L_i } \;.
\end{equation}
We refer to $\sigma_{z,l}$ and~$v_{z,l}$ weighted by the bolometric
luminosities as $\sigma_{z,l,bol}$ and~$v_{z,l,bol}$, respectively, and
those weighted by the Fe--K luminosities as $\sigma_{z,l,{\rm Fe}}$
and~$v_{z,l,{\rm Fe}}$, respectively. The velocities are shown in
Figure~\ref{fig:vel} and $\sigma_{z,l,{\rm Fe}}$
and~$v_{z,l,{\rm Fe}}$ are also shown in Table~\ref{tab:para}.

\section{Discussion}
\label{sec:dis}

In Figure~\ref{fig:sp}, the line profiles appear nearly symmetric.  This
is because the probability distribution function for $v_z$ is almost
symmetric (Matsumoto et al. 2004, in preparation).  Recently,
\citet{ino03} investigated line emission from turbulent gas in detail
and indicated that line profiles could be very complicated; asymmetry
would be observed in the X-ray spectra. However, for the turbulence we
considered, the expected asymmetry is too small to be detected with the
energy resolution of {\it Astro-E2 XRS}.  On the other hand, the
asymmetric injection of tsunamis is observed as a large offset of
$v_z\sim -140\:\rm km\: s^{-1}$ (Model~B; Table~\ref{tab:para}), because
the asymmetric tsunamis (e.g. those created by cluster mergers)
sometimes induce not only the turbulence but also the bulk motion (or
oscillation) of the cool core; the latter is unlikely to be induced by
symmetric jets accompanied by AGN activities. Therefore, the relative
motion between the cD galaxy (stars) and the strongly turbulent core
(gas) in a cluster could be a clue to confirm the tsunami model as well
as the detection of turbulence in clusters without AGN activities.

\begin{figure}
\epsscale{.45} 
\plotone{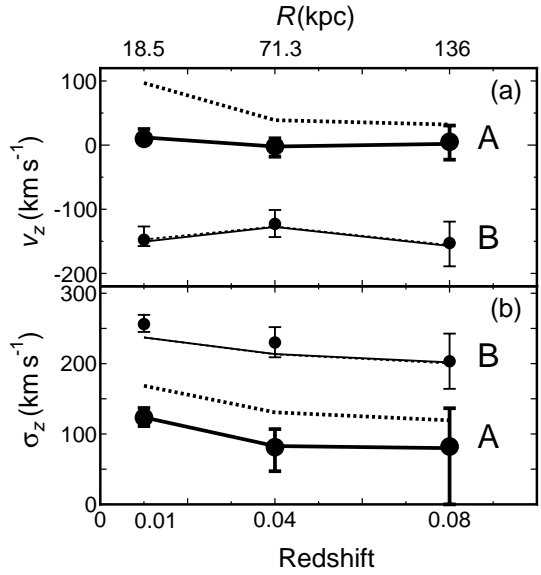} 
\caption{(a) Average gas velocities for different redshifts. Radii corresponding to the field of view of {\it XRS} ($R$) are also shown. Dotted lines indicate$v_{z,l,bol}$ andsolid lines indicate $v_{z,l,{\rm Fe}}$. Filled circles indicate $v_{z,sp}$. Bold lines and large circles correspond to Model~A and thin lines and small circles correspond to Model~B. (b) Turbulent velocities for different redshifts. Dotted lines indicate $\sigma_{z,l,bol}$ and solid lines indicate
$\sigma_{z,l,{\rm Fe}}$. Filled circles indicate $\sigma_{z,sp}$. Bold lines and large circles correspond to Model~A and thin lines and small circles correspond to Model~B.
\label{fig:vel}}
\end{figure}

Figure~\ref{fig:vel} shows that $v_{z,sp}$ and $v_{z,l,{\rm Fe}}$ are
consistent, and also show that $\sigma_{z,sp}$ and $\sigma_{z,l,{\rm
Fe}}$ are consistent. These mean that velocity fields obtained through
hydrodynamic simulations can directly be compared with X-ray
observations without resorting to mock observations like the one we did
here, if the velocities are weighted by line-emission. On the other
hand, near the cluster center, which means at smaller redshift, in
Model~A, $v_{z,sp}$ and $\sigma_{z,sp}$ are not consistent with
$v_{z,l,bol}$ and $\sigma_{z,l,bol}$, respectively
(Figure~\ref{fig:vel}). However, they are consistent in Model~B. The
difference between Model~A and~B is because radiative cooling proceeds
more and the gas temperature at the cluster center is lower for Model~A
at $t=3.3$~Gyr than for Model~B at $t=5.0$~Gyr. In Model~A, the
temperature near the cluster center is $< 3$~keV, which is much smaller
than the temperature outside of the core ($\sim 7$~keV) and the Fe--K
emission is relatively weak. Since cooler gas tends to have larger bulk
and turbulent velocities, the gases with larger velocities are less
weighted by the Fe--K luminosities.  Therefore, in Model~A, the
velocities weighted by the bolometric luminosities ($|v_{z,l,bol}|$ and
$\sigma_{z,l,bol}$) are larger than those weighted by the Fe--K
luminosities ($|v_{z,l,{\rm Fe}}|$ and $\sigma_{z,l,{\rm Fe}}$) near the
cluster center.  On the other hand, in model~B, the gas temperature in
the central region of the cluster is $\sim 4$~keV and the temperature
gradient near the center is small compared to that in
Model~A. Therefore, $v_{z,l,bol}$ ($\sigma_{z,l,bol}$) and $v_{z,l,{\rm
Fe}}$ ($\sigma_{z,l,{\rm Fe}}$) are almost the same.

In Figure~\ref{fig:vel}, the luminosity-weighted turbulent velocities,
$\sigma_{z,l,bol}$ and $\sigma_{z,l,{\rm Fe}}$,
increase toward the cluster center (toward smaller redshift). This means
that the turbulence produced by tsunamis is more detectable at the
cluster center.

As mentioned in \S\ref{sec:const}, we used a low resolution
grid. However, we expect that the effect of the coarsening on the
$v_{z,sp}$ and $\sigma_{z,sp}$ is small, because
$v_{z,l,{\rm Fe}}$ and
$\sigma_{z,l,{\rm Fe}}$ are not much different in
between the low resolution grid and the highest one; the difference is
much smaller than the errors of $v_{z,sp}$ and $\sigma_{z,sp}$ presented
in Table~\ref{tab:para} and Figure~\ref{fig:vel}. Since $v_{z,sp}$ and
$v_{z,l,{\rm Fe}}$ ($\sigma_{z,sp}$ and
$\sigma_{z,l,{\rm Fe}}$) are almost same as discussed
above, $v_{z,sp}$ ($\sigma_{z,sp}$) will not change much even for the
finest grid.

We also `observed' the cluster from the direction perpendicular to the
$z$-axis and we call this direction $x$. Because of the symmetry we
assumed, the average velocities, $v_x$, are zero. The turbulent
velocities derived from the mock observations are $\sigma_{x,sp}\approx
60$ and $120~\rm\: km \:s^{-1}$ for Model~A and~B,
respectively. Consistency between $\sigma_{x,sp}$ and those weighted by
Fe--K lines, $\sigma_{x,l,Fe}$, is also good.

\section{Conclusions}
\label{sec:conc}

We have constructed X-ray spectra from the results of hydrodynamic
simulations of clusters of galaxies based on the tsunami model, in which
turbulence is created in cluster cores. Similar levels of turbulence are
also predicted by many of other heating models of cool cores
\citep[e.g. motion of bubbles created by AGN
activities;][]{chu02,bru02}. In particular, we focus on the effect of
velocity fields of the ICM on the spectra. We simulate X-ray
observations with {\it Astro-E2 XRS} and find that velocity fields in
cluster cores could be revealed with the satellite. The motion of the
cool core could be used to discriminate among turbulence models We show
that the gas velocities derived through the mock observations are
consistent with the Fe--K line emission weighted values inferred
directly from hydrodynamic simulations. The technique developed here
could easily be applied to the comparison between results of various
hydrodynamic simulations and those of near-future observations with {\it
Astro-E2} and others ({\it Constellation-X, XEUS, NeXT}).

\acknowledgments

We thank the anonymous referee for helpful suggestions.  We thank
N. Aghanim, H. Matsumoto and M, Sakano for useful discussion. We are
grateful to T. Hanawa for contribution to construction of the nested
grid code. The authors are supported in part by a Grant-in-Aid from the
Ministry of Education, Culture, Sports, Science, and Technology of Japan
(Y. F.: 14740175; T. M.: 16740115; K. W.: 15684003). T. F. is supported
by the Japan Society for the Promotion of Science.

\clearpage

\begin{deluxetable}{cccc}
\tabletypesize{\scriptsize}
\tablecaption{Model Parameters}
\tablewidth{0pt}
\tablehead{
\colhead{Model} & \colhead{$\alpha$} & \colhead{$\lambda$ (kpc)} 
& \colhead{$t$ (Gyr)}
}
\startdata
A & 0.3 & 100 & 3.3 \\
B & 0.3 & 1000 & 5.0 \\
\enddata
\label{tab:model}
\end{deluxetable}

\clearpage

\begin{deluxetable}{ccccccccc}
\tabletypesize{\scriptsize}
\tablecaption{Fitting Results}
\tablewidth{0pt}
\tablehead{
\colhead{Model}  & \colhead{$T$} & \colhead{$Z$} 
& \colhead{$v_{z,sp}$} & \colhead{$\sigma_{z,sp}$}
& \colhead{$\chi^2/$dof} 
& \colhead{$v_{z,l,{\rm Fe}}$\tablenotemark{a}} 
& \colhead{$\sigma_{z,l,{\rm Fe}}$\tablenotemark{b}}
& \colhead{$F_x$ (5--10~keV)\tablenotemark{c}} 
\\
\colhead{(Redshift)}  & \colhead{(keV)} & \colhead{($Z_\sun$)} 
& \colhead{($\rm km\: s^{-1}$)} & \colhead{($\rm km\: s^{-1}$)} 
& \colhead{} 
& \colhead{($\rm km\: s^{-1}$)} & \colhead{($\rm km\: s^{-1}$)}
& \colhead{($\:\rm erg\: cm^{-2}\: s^{-1}$)} 
}
\startdata
 A (0.01) & $2.80_{-0.05}^{+0.06}$ & $0.26_{-0.01}^{+0.01}$ 
& $10_{-5}^{+15}$ & $123_{-12}^{+14}$
& 217.6/679 
& 12 & 124 
& $1.1\times 10^{-10}$ \\
 A (0.04) & $4.55_{-0.19}^{+0.20}$ & $0.28_{-0.02}^{+0.02}$ 
& $-2_{-16}^{+13}$ & $81_{-34}^{+26}$
& 122.8/241
& -2 & 83
& $3.2\times 10^{-11}$ \\
 A (0.08) & $5.71_{-0.47}^{+0.41}$ & $0.30_{-0.03}^{+0.03}$ 
& $5_{-28}^{+25}$ & $82_{-82}^{+54}$
& 170.4/95
& 2 & 80
& $1.2\times 10^{-11}$ \\
 B (0.01) & $4.04_{-0.09}^{+0.11}$ & $0.29_{-0.01}^{+0.01}$ 
& $-147_{-10}^{+21}$ & $256_{-11}^{+13}$
& 163.8/724
& -150 & 237
& $1.1\times 10^{-10}$ \\
 B (0.04) & $4.44_{-0.19}^{+0.17}$ & $0.30_{-0.02}^{+0.02}$ 
& $-123_{-21}^{+22}$ & $230_{-21}^{+22}$
& 128.5/283
& -128 & 214
& $3.8\times 10^{-11}$ \\
 B (0.08) & $4.81_{-0.33}^{+0.36}$ & $0.29_{-0.03}^{+0.03}$ 
& $-153_{-36}^{+33}$ & $203_{-39}^{+39}$
& 88.2/104
& -158 & 201
& $1.4\times 10^{-11}$ \\
\enddata
\tablenotetext{a}{Average velocity weighted by the Fe--K lines
 luminosities}
\tablenotetext{b}{Turbulent velocity weighted by the Fe--K lines
 luminosities}
\tablenotetext{c}{X-ray flux in the 5--10~keV band}
\label{tab:para}
\end{deluxetable}

\clearpage

%\begin{figure}
%\epsscale{.80} 
%\plotone{f1.eps} 
%\caption{The temperature distribution of a cluster core for Model~B at $t=5.0$~Gyr. (a) for $z\lesssim 300$~kpc, (b) for $z\lesssim 40$~kpc. Note that color bars are different between (a) and (b).  
%\label{fig:temp}}
%\end{figure}

\clearpage

\clearpage

\end{document}